\documentclass{emulateapj} 
\usepackage{apjfonts,psfig} 
\begin{document}
\slugcomment{ApJ, in press}
\shortauthors{M. Catelan} 
\shorttitle{RR Lyrae Variables in M3: Observations and Theory}

\title{The Evolutionary Status of M3 RR Lyrae Variables: \\
       Breakdown of the Canonical Framework?}

  \author{M.~Catelan} 
   \affil{Pontificia Universidad Cat\'olica de Chile, Departamento de 
Astronom\'\i a y Astrof\'\i sica, \\ Av. Vicu\~{n}a Mackenna 4860, 
782-0436 Macul, Santiago, Chile; email: mcatelan@astro.puc.cl}

\begin{abstract}
In order to test the prevailing paradigm of horizontal-branch (HB) stellar evolution, 
we use the large databases of measured RR Lyrae parameters for the globular cluster  
M3 (NGC~5272) recently provided by Bakos et al. and Corwin \& Carney. We compare the 
observed distribution of fundamentalized periods against the predictions of synthetic 
HBs. The observed distribution shows a sharp peak at $P_{\rm f} \approx 0.55$~d, which 
is primarily due to the RRab variables, 
whereas the model predictions instead indicate that the distribution should be more 
uniform in $P_{\rm f}$, with a buildup of variables with shorter periods 
($P_{\rm f} < 0.5$~d). Detailed statistical tests show, for the first time, that the 
observed and predicted distributions are incompatible with one another at a high 
significance level. Either this indicates that canonical HB models are inappropriate, 
or that M3 is a pathological case that cannot be considered representative of the 
Oosterhoff type I (OoI) class. In this sense, we show that the OoI cluster with 
the next largest number of RR Lyrae variables, M5 (NGC~5904), presents a similar, 
though less dramatic, challenge to the models. We show that the sharp peak in the 
M3 period distribution receives a significant contribution from the Blazhko variables 
in the cluster. We also show that M15 (NGC~7078) and M68 (NGC~4590) 
show similar peaks in their $P_{\rm f}$  
distributions, which in spite of being located at a similar $P_{\rm f}$ value as 
M3's, can however be primarily ascribed to the RRc variables. Again similar to M3, 
a demise of RRc variables towards the blue edge of the instability strip is also 
present in these two globulars. This is again in sharp contrast with the 
evolutionary scenario, which also foresees a strong buildup of RRc variables with 
short periods in OoII globulars. We speculate that, in OoI systems, RRab variables 
may somehow get ``trapped'' close to the transition line between RRab and RRc pulsators 
as they evolve to the blue in the H-R diagram, whereas in OoII systems it is the RRc 
variables that may get similarly ``trapped'' instead, as they evolve to the red, 
before changing their pulsation mode to RRab. Such a scenario is supported by the 
available CMDs and Bailey diagrams for M3, M15, and M68. 
\end{abstract}


\section{Introduction}
For many years, M3 (NGC~5272) has been considered the canonical globular cluster 
(GC) par excellence. Lowly reddened, relatively nearby, and having a long 
history of detailed observations, M3 was early noted to contain hundreds of RR 
Lyrae (RRL) variables, making it a ``splendid object for further studies'' 
(Bailey 1902). This, of course, was confirmed by  
all subsequent analyses (Bakos, Benk\"{o}, \& Jurcsik 2000; Corwin \& Carney 
2001; Clement et al. 2001 and references therein), which revealed that M3 is 
the GC with the largest (known) number of RRL 
variables, apart from $\omega$~Centauri (NGC~5139). Hence it is not surprising 
to find that M3 is the prototype of the so-called Oosterhoff type I (OoI) 
variability class (Oosterhoff 1939). For comparison purposes, the other OoI 
cluster noted by Oosterhoff, M5 (NGC~5904)---also the OoI GC with the second 
largest number of RRL variables---contains $\sim 100$ fewer 
known RRL variables than M3. 

In the present paper, we take benefit of the 
uniquely large number of RRL variables in M3 to perform, for the first time, a 
detailed statistical comparison between the predictions of the standard model 
of horizontal-branch (HB) evolution and RRL pulsation and the observed periods. 
In particular, our aim is to investigate whether the discrepancy between 
predicted and observed period distributions originally noted by Rood \& Crocker 
(1989) is confirmed by the current models. In \S2, we describe our analysis 
techniques. In \S3, we compare the RRL variables in the inner and outer regions 
of M3. In \S4, we describe our reference M3 models, which are then compared 
against the observations. In \S6, we discuss the impact of different assumptions 
about the instability strip (IS) topology upon our results. Finally, in \S7, we 
discuss possible explanations for the discrepancies we find, and extend our 
analysis to the cases of other GCs.

\section{Analysis}
We retrieved and employed the catalogues of RRL periods, photometric 
properties and coordinates from Bakos et al. 
(2000)\footnote{http://www.konkoly.hu/staff/bakos/M3/table.html.} 
and Corwin \& Carney (2001). These contain a total of $>200$ RRL  
with determined periods that can be used in our analysis. 
Info from these catalogues includes 
periods, variability type and mode of pulsation, information on the 
presence (or otherwise) of the Blazhko effect, colors, magnitudes, 
and radial distance from the center of the cluster. At the outset, 
we note that the mean period of the fundamental-mode pulsators (RRab's), 
$\langle P_{\rm ab} \rangle = 0.559 \pm 0.005$~d, as well as the 
number fraction of first-overtone pulsators (RRc's), $22\pm 2\%$, 
confirm Oosterhoff's (1939) definition of the OoI group based on M3. 

The simulations employed in the present paper are similar to those described 
in Catelan, Ferraro, \& Rood (2001). In particular, the evolutionary tracks 
are the same as those computed by Catelan et al. (1998), and assume a main 
sequence chemical composition $Y_{\rm MS} = 0.23$, $Z = 0.001$; the HB stars 
have an envelope helium abundance $Y_{\rm HB} = 0.2409$, due to the 
extra helium brought to the surface during the first dredge-up. 
The HB synthesis code, {\sc sintdelphi}, is an 
updated version of Catelan's (1993) code. To compute a synthetic HB, the 
code assumes that the mass distribution on the zero-age HB (ZAHB) is a 
normal deviate (Rood 1973; Caputo et al. 1987; Lee 1990; 
Catelan et al. 1998). This standard 
assumption is now known to break down in the cases of at least some of 
the so-called ``bimodal HB clusters'' 
(e.g., Rood et al. 1993; Catelan et al. 1998, 2002), as well as 
in clusters with significant populations of ``extreme HB stars'' (e.g., D'Cruz 
et al. 1996)---but has generally been considered a reasonable approximation 
for clusters with ``well-behaved'' HBs such as the ones we model in the 
present paper (e.g., Lee 1990), and even in clusters with long blue tails 
such as M79 (Dixon et al. 1996). Note, in addition, that the adoption of 
normal {\em deviates}, as opposed to strictly normal distributions, along 
with the fact that the HB synthesis technique adopts a Monte Carlo approach, 
implies that individual cases drawn from large pools of simulations will 
often bear little resemblance to an actual Gaussian distribution. 

By default, the blue edge (BE) of the IS is computed for each individual star 
as a function of its basic physical parameters ($L$, $M$, $Y_{\rm HB}$) 
using eq.~(1) of Caputo et al. (1987), which is based on the Stellingwerf 
(1984) pulsation models. Then, for each star, the temperature 
of the red edge (RE) of the IS is obtained based on a free 
input parameter, the width of the IS $\Delta\,\log T_{\rm eff}^{\rm IS}$, 
that is supplied at run time. If the temperature of the star falls in between 
the so-computed blue and red edges of the IS, it is then flagged 
as an RRL variable, and its (fundamental) pulsation period computed 
according to eq.~(4) in Caputo, Marconi, \& Santolamazza (1998)---an updated 
version of the van Albada \& Baker (1971) period-mean density relation. 
In \S6, we will discuss the implications of alternative formulations for 
the IS edges upon our results. 

In obtaining the ``best-fitting'' HB simulations, we start from the solutions 
provided by Catelan et al. (2001), in the case of M3 (see \S\S\S4,5,6), and 
by Catelan (2000), in the case of M5 (see \S7); the reader is referred to 
these papers for an assessment of the quality of the fits. In the present 
paper, whenever changes in the input parameters (mean mass on the ZAHB and 
mass dispersion) are required in comparison with those studies---due, for 
instance, to changes in the placement of the BE of the IS---we 
make sure, in particular, that these ``perturbed'' solutions, 
which in general differ only slightly from the original ones, also provide 
excellent matches to the observed number counts along the cluster HBs, i.e., 
$B,V,R/(B+V+R)$.

\section{Inner vs. Outer Regions}
Given the recent discovery (Catelan et al. 2001; Catelan, Rood, \& 
Ferraro 2002) that the HB morphology of M3 is significantly bluer in the 
innermost regions ($r < 50\arcsec$) than in the outermost 
ones ($r > 210 \arcsec$), and predictions, based on stellar 
evolution theory and the period-mean density relation, that the pulsation 
periods of RRL stars may depend on the HB type (e.g., Lee, Demarque, \& 
Zinn 1990), we have initially investigated whether the period distribution of 
RRL variables may differ between the two noted radial regions. In order 
to perform the test, we have followed the recommendations of van Albada \& 
Baker (1973), ``fundamentalizing'' the RRc 
variables by adding 0.128 to the logarithm of their periods. 
We then ran a Student $t$-test to check whether the distributions of 
fundamentalized periods over the two noted radial regions are different, 
finding instead a $\simeq$~47\% probability that the two 
distributions are derived from the same parent distribution. 
This is confirmed by a two-sample Kolmogorov-Smirnov (K-S) test, which implies 
that the two distributions are drawn from the same parent distribution with 
84.5\% probability. 

We thus conclude that the range in HB type within 
M3, though significant, is insufficient to affect the pulsation properties 
of its variables, so that it is safe to employ the complete sample of M3 
variables in our tests. This conclusion is fully supported by HB simulations 
independently computed for the inner and outer regions of the cluster.

\section{Global Analysis}\label{glob}
Extending thus the HB number counts reported in Catelan et al. 
(2001) over all radial regions of M3 (F. R. Ferraro 2000, priv. comm.), 
we obtain $B+V+R = 530$,  
$B:V:R = 0.39: 0.40: 0.21$, and $(B-R)/(B+V+R) = + 0.183$. The best-fitting, 
canonical, unimodal HB simulation that best reproduces these parameters, 
under the prescription for the BE of the IS discussed 
in $\S2$, and adopting for the width of the RRL strip the canonical 
value $\Delta\log T_{\rm eff}^{\rm IS} = 0.085$ (Smith 1995, \S1.2.3 and Table~1.1), 
is characterized by a mean ZAHB mass value and a mass 
dispersion  given by, respectively, 

\begin{displaymath}
\langle M_{\rm HB} \rangle = 0.641 \, M_{\odot}, \,\,\,\,\, 
\sigma_M = 0.019 \, M_{\odot}. \,\,\,\,\, \,\,\,\,\, \,\,\,\,\, {\rm (Case \,\, A)}
\end{displaymath}

\noindent Not surprisingly, both of these parameters are intermediate between 
the values for the ``inner'' and ``outer'' solutions given in Catelan et al.
(2001). These are the parameters that we shall use in \S5 
(``Case A''). Cases B and C will be described in \S6.

\begin{figure*}[t]
  \figurenum{1}
  \centerline{\psfig{figure=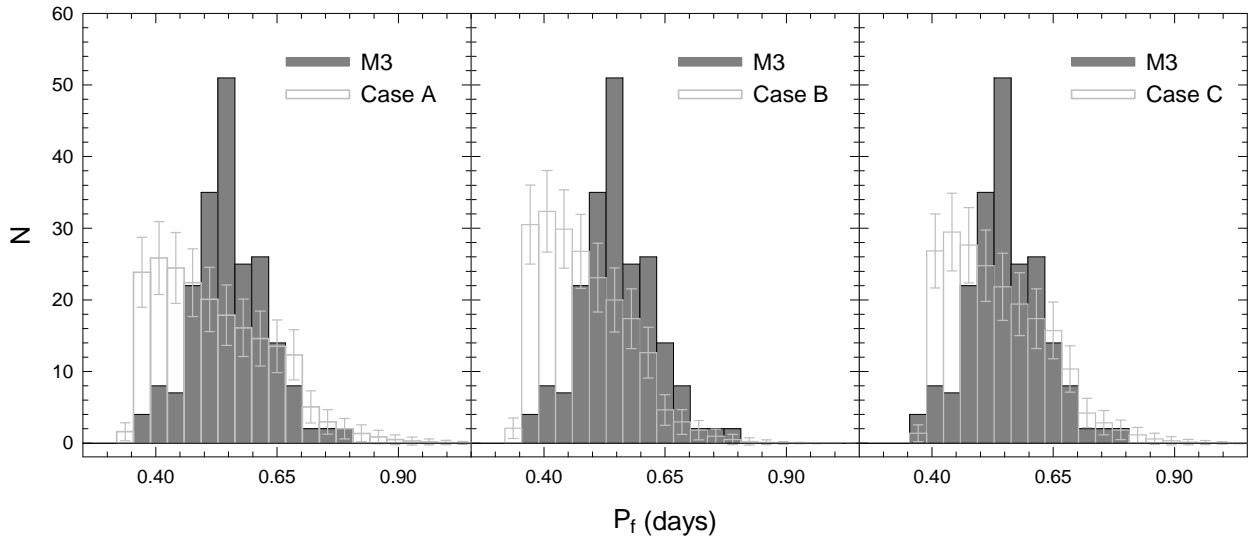,width=6.5in}}
  \caption{Histogram of fundamentalized periods of M3 RRL variables: 
     Observations: filled bars; normalized simulations: 
     light gray bars, with Poissonian errors indicated. 
      }
      \label{Fig01}
\end{figure*}

\begin{figure*}[t]
  \figurenum{2}
  \centerline{\psfig{figure=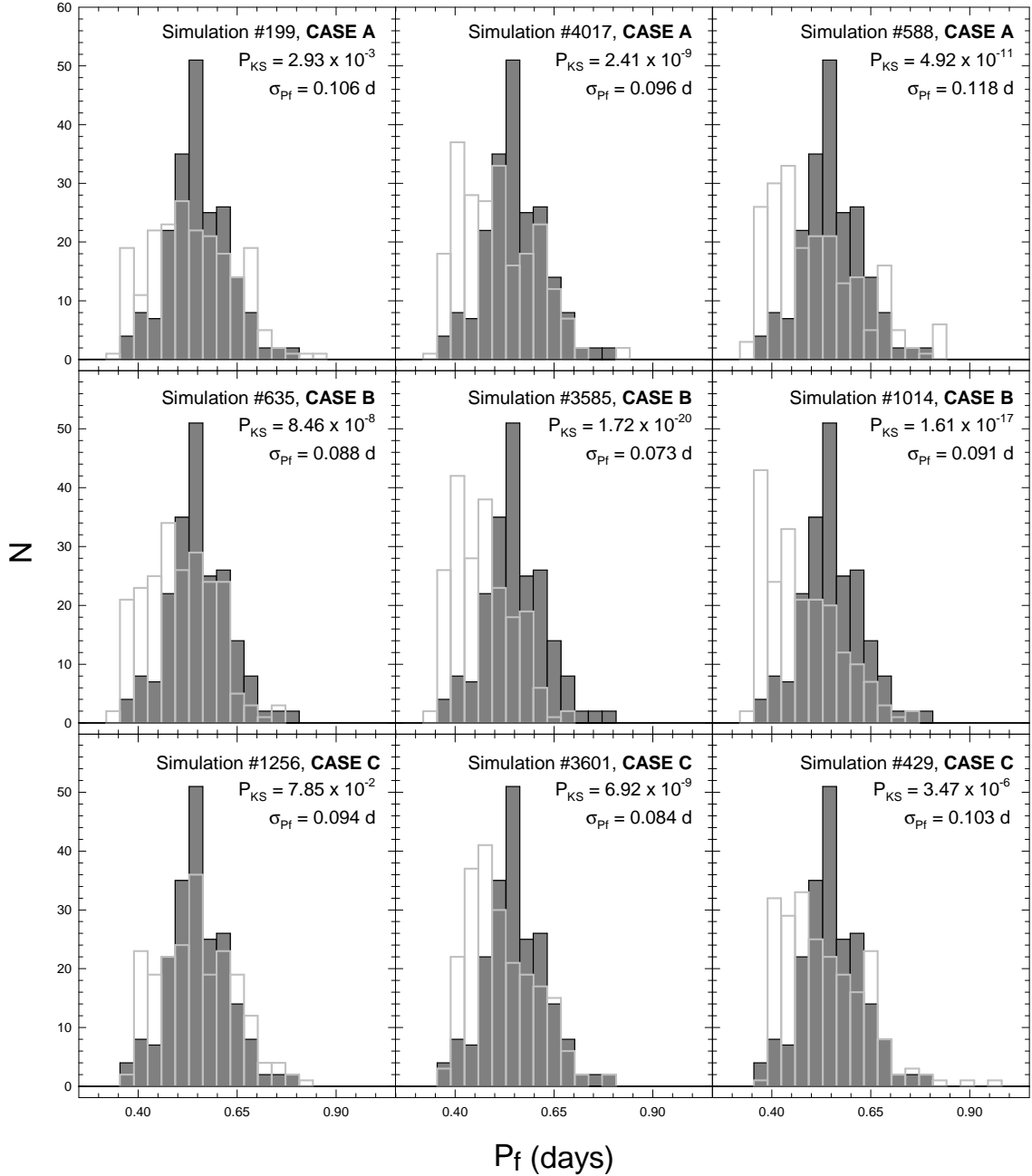,width=5.92in}}
  \caption{Period distributions for synthetic HB models for M3 (light 
        gray bars), compared with the observations (filled bars). 
        Top row: ``Case~A'' models, which assume an IS 
        width of $\Delta\log T_{\rm eff}^{\rm IS} = 0.085$ and have a 
        blue edge of the IS at $T_{\rm eff} \simeq 7400$~K. 
        Middle row: ``Case~B'' models, which are similar 
        to Case~A models but assume a narrower IS width, 
        $\Delta\log T_{\rm eff}^{\rm IS} = 0.070$. 
        Bottom row: ``Case~C''
        models, which assume the same IS width as in Case~B, but  
        whose blue edge of the IS is cooler by 200~K. 
        For each case, we have, from left to right: the 
        model with the highest $P_{\rm KS}$; the model with the smallest 
        $\sigma_{P_{\rm f}}$; and a randomly picked model. }
      \label{Fig02}
\end{figure*}

\begin{figure*}[t]
  \figurenum{3}
  \centerline{\psfig{figure=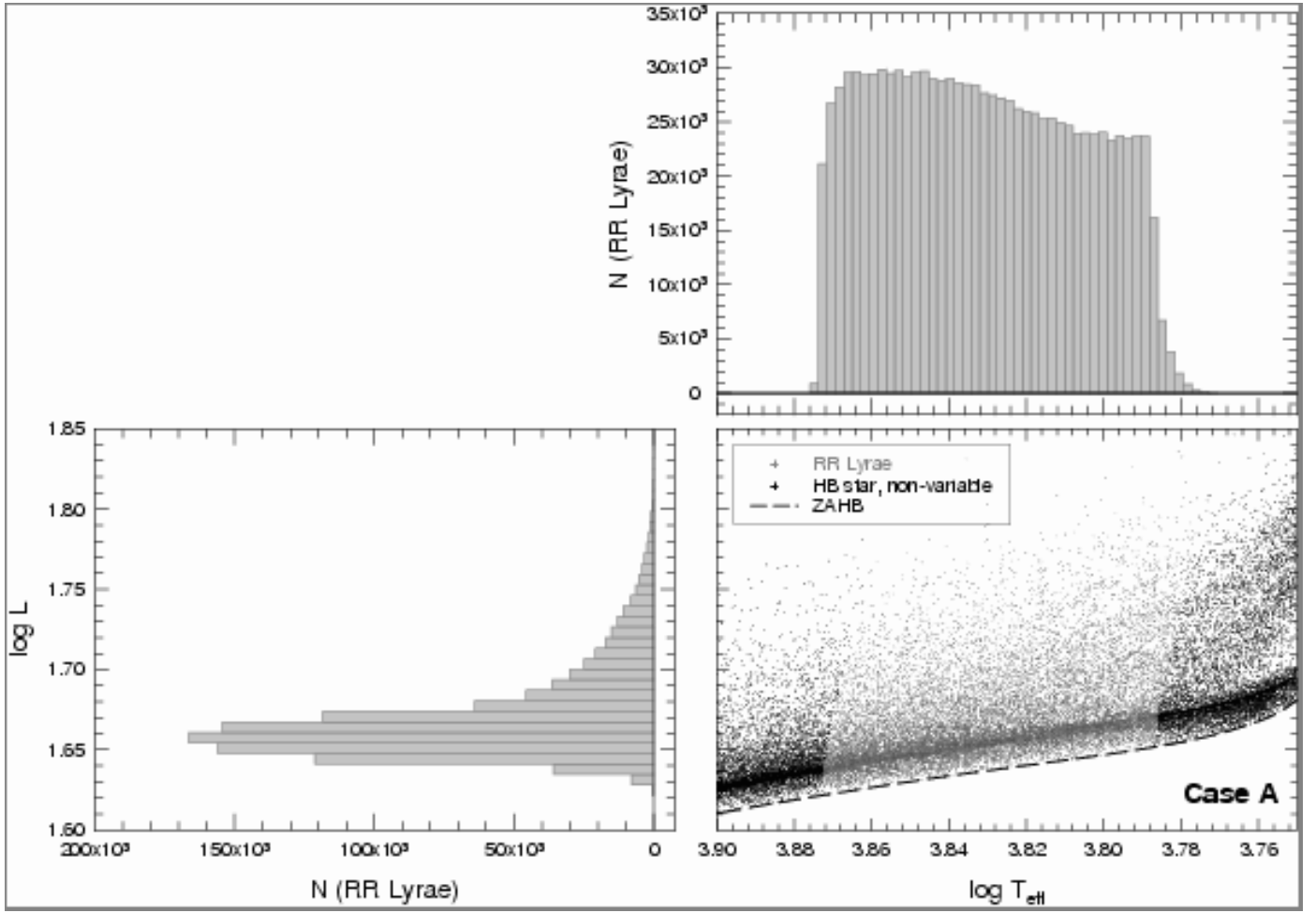,width=7.25in}}
  \caption{The H-R diagram at the bottom right, centered around the RRL region, 
     represents a subsample of 20,000 RRL stars from the complete pool 
     of 1,076,042 RRL stars in the 5000 simulations for Case~A. A 
     proportional number of non-variable HB stars is also shown. The 
     temperature and luminosity histograms (top and left panels, 
     respectively) refer to the full set of RRL stars from the 
     Case~A simulations. In the H-R diagram, the ZAHB line is 
     indicated. Note the fairly uniform temperature distribution, 
     and the sharply peaked luminosity distribution 
     at a luminosity level well {\em above} the ZAHB.} 
      \label{Fig03}
\end{figure*}

\begin{figure}[t]
  \figurenum{4}
  \centerline{\psfig{figure=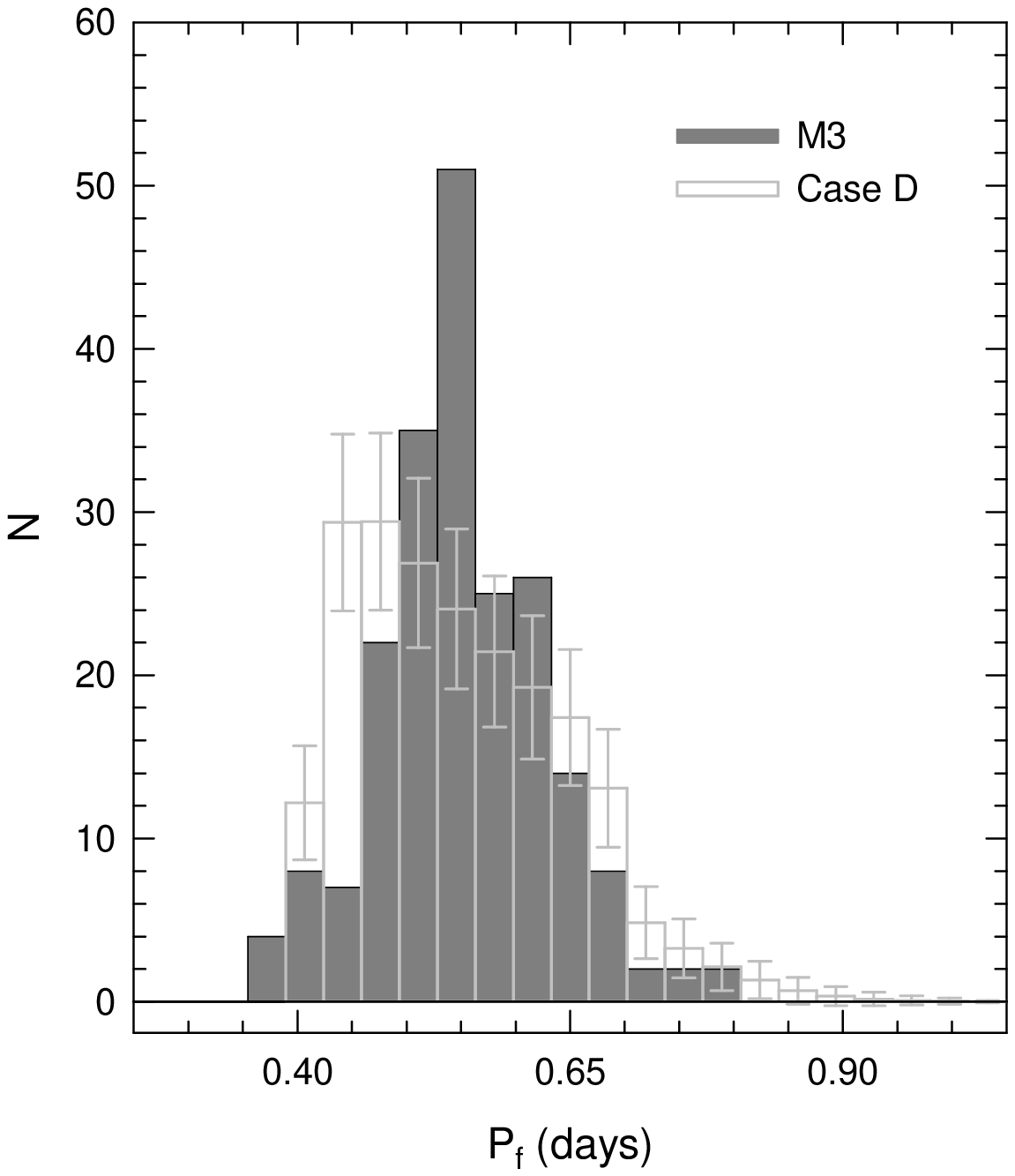,width=3.25in}}
  \caption{As in Fig.~1, but assuming a $\Delta\log T_{\rm eff}^{\rm IS} = 0.065$, 
    and shifting the BE by $-300$~K with respect to Case~A. 
      }
      \label{Fig04}
\end{figure}

\section{Theory versus Observations} 
We first compare the predicted and observed (fundamentalized) period 
distributions using a two-sample K-S test. A set of 5000 simulations with 
the input parameters described in \S4 (``Case~A'') yields a mean  
probability that the computed and observed periods are drawn from the same 
parent distribution of 
$\langle P_{\rm KS}\rangle \approx 4.5 \times 10^{-6}$. The highest value of 
$P_{\rm KS}$, in these simulations, was 0.29\%.  
Thus, the simulated and 
observed period distributions are different at a high significance 
level. Following Rood \& Crocker (1989), we attribute this to the peaked 
shape of the M3 distribution, which is shown in Fig.~1, left panel, along 
with the co-added, normalized simulations. We have checked that the comparison 
between models and observations, as shown in Fig.~1, remains qualitatively 
unchanged when the number of bins is allowed to vary in the range from 15 to 
25, with the bin size automatically computed from the number of bins and the 
plot limits 0.25~d and 1.05~d. Note that our statistical analysis is 
{\em insensitive} to the binning, since it relies on a K-S test; the histograms 
are shown only for illustration purposes, and to highlight the period ranges 
over which the discrepancy between models and observations is most severe 
(see \S6 for a more detailed discussion).  

In order to further test the frequency of ``peaked'' period distributions 
that might resemble M3's, we have also computed the standard deviation of the 
fundamentalized period distribution 
for all the simulations. This could be relevant, in particular, if 
peaked distributions occurred somewhat randomly in $P_{\rm f}$ space, 
the precise location of the M3 peak not bearing particular significance. 
We find 
$\langle \sigma_{P_{\rm f}} \rangle = 0.116\pm 0.006$~d (standard deviation), 
whereas the observed value is 
$\sigma_{P_{\rm f}} = 0.077$~d. Thus the observed distribution 
is intrinsically more ``peaky'' than the simulated ones---a $6\sigma$ 
result. Clearly, Case~A simulations cannot provide a good 
description of M3 RRL properties. 

We provide the 
histograms for three simulations computed in this way in the top row of 
Fig.~2 (``Case~A''), where the models with the largest $P_{\rm KS}$ and 
smallest $\sigma_{P_{\rm f}}$ are given 
in the left and middle panels, respectively. 
The right panel in that row corresponds to 
a randomly picked simulation from the remaining pool of 4998 
synthetic HBs. Note that the axis scales and bin sizes are
identical to the ones used in Fig.~1. 

Analysis of these plots discloses that the theoretical 
models, besides presenting much less sharp peaks than observed, 
also present a build-up of stars at short periods which is not matched by  
the observed distribution. This is due to the fact 
that the RRc side of the IS is as well populated as the RRab 
side, as can be seen by the H-R diagram and $\log T_{\rm eff}$ histogram 
for the co-added Case~A simulations (Fig.~3, bottom right and top panels), 
and also to the fact that relatively fainter regions of the H-R diagram 
tend to be more populated (due to slower evolution) than the brighter 
parts of the IS at a given temperature (Fig.~3, bottom panels). The signature 
of these brighter stars can be identified in the long-period ``tail'' 
that is seen in the theoretical period histograms; however, the drop from 
the peak to this long-period tail is more abrupt in the data, which is also 
reflected upon the systematically smaller observed $\sigma_{P_{\rm f}}$ in 
comparison with the predicted values. Note that qualitatively similar 
features had previously been found by Rood \& Crocker (1989), whose 
results are fully supported by our simulations. 

We call attention, in passing, to the fact that the ZAHB strictly corresponds 
to a lower, fairly thinly populated lower 
envelope to the distribution in the H-R diagram 
(Fig.~3), the bulk of the HB stars being found at slightly larger luminosities, 
while the HB stars are evolving on blue loops. Such a phenomenon was obvious 
in previous studies as well---see, e.g., Fig.~4b in Catelan (1993), Fig.~17 
in Catelan et al. (1998), 
and also Ferraro et al. (1999). For this reason, it appears unlikely 
to us that the stars that tend to clump around the ``OoI line'' in the 
Bailey diagram are actual ``ZAHB stars,'' as suggested by Clement \& Hazen 
(1999). We suggest instead that many of these stars are more likely to be ``blue 
loop stars.'' Also in passing, we note that the RRL which have been found ``below 
the ZAHB'' in M3 (Corwin \& Carney 2001; see also Jurcsik 2003) may simply 
be a consequence of an imperfect analogy between zero-age MS, on the one  
hand---a line very close to which low-mass MS stars will indeed spend most of 
their MS lifetimes---and ZAHB, on the other---the ZAHB {\em not} being, in 
general, a line very close to which RRL/HB stars will spend most of their HB 
lifetimes.

\section{Instability Strip Edges}
The results of Catelan et al. (2001) indicate that the IS boundaries, 
calculated as described in \S2, provide a good description of the M3 IS 
boundaries (see Fig.~1 in Catelan et al.). The color transformations adopted 
in Catelan et al. correspond to an earlier version of the VandenBerg \& Clem 
(2003) transformations, which were kindly provided by D. A. VandenBerg (1999, 
priv. comm.) prior to publication. Note that the (average) temperature of the 
BE of the IS, as computed in \S5 following the prescriptions 
described in \S2, is $\approx 7420$~K, thus being similar to the canonical 
value reported in Smith (1995).  

However, Smith (1995) also cautions that such a BE may be too hot, 
``perhaps by 100--200~K''---and current pulsation models do seem to favor 
a slighly cooler BE for the relevant metallicities ($Z \approx 0.001-0.002$), 
as can be seen from Table~1 in Caputo et al. (2000; see also Bono et al. 1997). 
It is interesting to note 
that the quoted Caputo et al. pulsation models, for parameters similar to those 
implied by our simulations for the M3 RRL variables---namely, mean mass 
$\langle M_{\rm RR} \rangle \simeq 0.645 \, M_{\odot}$, mean luminosity 
$\langle \log\,(L_{\rm RR}/L_{\odot}) \rangle \simeq 1.67$---imply  
an IS width that is even larger than what we assumed in \S5, namely, 
$\Delta\log T_{\rm eff}^{\rm IS} = 0.086-0.089$. The BE of the IS is found, according 
to these models, at about $T_{\rm eff} \simeq 7150-7250$~K; this is consistent with 
Smith's remark about a possible shift of the IS towards lower temperatures. 
Moreover, it appears reasonable to assume that the RE of the IS---hence 
the IS width, once the BE temperature is fixed---remains more uncertain than the 
BE at the present time, in spite of the efforts that were made in the quoted 
papers to take into account nonlocal and time-dependent convective effects in 
the nonlinear computations (see also Feuchtinger 1999)---ingredients which, along
with the choice of mixing-length parameter, do not play as significant a role in 
the case of the IS BE. 
Moreover, the 
conversion between colors and temperatures is still not without uncertainties 
at the RRL level---see, e.g, the comparison among predictions by different 
authors in Figs.~3, 4, and 19 of VandenBerg \& Clem (2003), bearing in mind the 
possible suggested ranges in RRL temperatures, from $\sim 5800$~K (Caputo et 
al. 2000, their Table~1) to $\sim 7400$~K (Smith 1995). 
For these reasons, 
at the present time, we cannot rule out the possibility that the IS 
is significantly narrower than we have assumed thus far. In fact, Fig.~7 
in Popielski, Dziembowski, \& Cassisi (2000), which seems to be based on a 
combination of theoretical models for the BE and empirical results for 
the RE, supports an IS width, at the luminosity level of the RRL in our 
simulations, of $\Delta\log T_{\rm eff}^{\rm IS} \simeq 0.079$, the width decreasing 
with decreasing luminosity. 

Therefore, in order to investigate the impact of IS topology uncertainties 
upon our results, we have also computed extensive sets of HB simulations for 
the following additional situations: ``Case~B'' has an IS width reduced to 
$\Delta\log T_{\rm eff}^{\rm IS} = 0.070$; ``Case~C'' not only has a narrower strip, 
but also a BE shifted by $-200$~K.\footnote{The referee notes that a new preprint 
has recently come out (Marconi et al. 2003) in which pulsation models are reported 
according to which the IS BE is indeed $226\pm 66$~K cooler than in 
our Case~A models.} 
For these cases, we now find that the best-fitting HB simulations are described 
by the following input parameters: 

\begin{displaymath}
\langle M_{\rm HB} \rangle = 0.640 \, M_{\odot}, \,\,\,\,\, 
\sigma_M = 0.015 \, M_{\odot}; \,\,\,\,\,\,\,\,\,\, \,\,\,\,\,  {\rm (Case \,\, B)} 
\end{displaymath}

\begin{displaymath}
\langle M_{\rm HB} \rangle = 0.643 \, M_{\odot}, \,\,\,\,\, 
\sigma_M = 0.015 \, M_{\odot}. \,\,\,\,\,\,\,\,\,\, \,\,\,\,\,  {\rm (Case \,\, C)}
\end{displaymath}

\noindent These are fairly similar to the Case~A solution for the mean ZAHB 
mass and mass dispersion (\S4), differences only appearing in the third decimal 
place for both quantities. However, the implications upon the predicted 
period distributions are more immediately apparent. 

Thus, for Case~B, we find $\langle P_{\rm KS}\rangle  \approx 4.4 \times 10^{-11}$, 
with the highest $P_{\rm KS}$ in the 5000 simulations being $8.5\times 10^{-8}$. 
Clearly, the discrepancy is even more significant in this case. 
We interpret this as being due to the necessity of producing synthetic HBs that 
match the observed number of RRL variables in M3: if the IS 
width is reduced, one must redistribute the variables that formerly 
fell on the low-temperature, longer-period end of the distribution to the 
other regions of the IS. Given that the highest probability 
for an RRL variable is to fall in the shorter-period, ``pile-up'' region 
(which is not seen in the data), that region is enhanced even further in this 
case, leading to an exacerbation of the problem. These effects are shown very 
clearly in the middle panel of Fig.~1. On the other hand, this case 
also gives a smaller $\langle \sigma_{P_{\rm f}} \rangle = 0.090\pm 0.005$~d, 
which however is still  
inconsistent with the observed $\sigma_{P_{\rm f}}$ at the $2.7\sigma$ level. 
The cases with the largest $P_{\rm KS}$ and smallest $\sigma_{P_{\rm f}}$, 
besides a randomly selected one, are given in the left, middle, and right panels, 
respectively, of the middle row of Fig.~2. We emphasize that, while some of 
these simulations do present reasonably sharp peaks, those peaks are invariably 
at the short-period end of the distribution, unlike the case in M3. Hence we 
conclude that Case~B simulations do not provide a good description of the M3 
period distribution either. 

Finally, Case~C provides 
$\langle P_{\rm KS}\rangle \simeq 7.21 \times 10^{-4}$ and 
$\langle \sigma_{P_{\rm f}} \rangle = 0.100\pm 0.005$~d. 
The highest $P_{\rm KS}$ found in the 5000 simulations for this case is $7.9\%$. 
Therefore, Case~C represents an improvement over the previous ones, though it 
is still far from being able to provide a satisfactory match to the observed 
distribution. The improvement chiefly results from the fact that, by moving 
the IS BE to a lower temperature, we are effectively ``forcing'' 
some of the RRL stars out of the ``pile-up region'' at the short-period end 
of the distribution---a feature which we have seen not to be present in the data. 
The flipside of the coin is that this also feeds the longer-period end of the 
distribution with more variables, resulting in a higher value of 
$\langle \sigma_{P_{\rm f}} \rangle$---in this case, inconsistent with the 
observed $\sigma_{P_{\rm f}}$ at the $4.4\sigma$ level. 
In those simulations where the period distribution does turn out 
to be peaky, such as in the middle panel in the bottom row of Fig.~2, 
the peak is still located towards the short-period end of the distribution, 
unlike the observed one which is primarily due to the RRab stars. Hence 
Case~C simulations cannot be considered satisfactory at explaining the shape of 
the M3 distribution.

\begin{figure*}[ht]
  \figurenum{5}
  \centerline{\psfig{figure=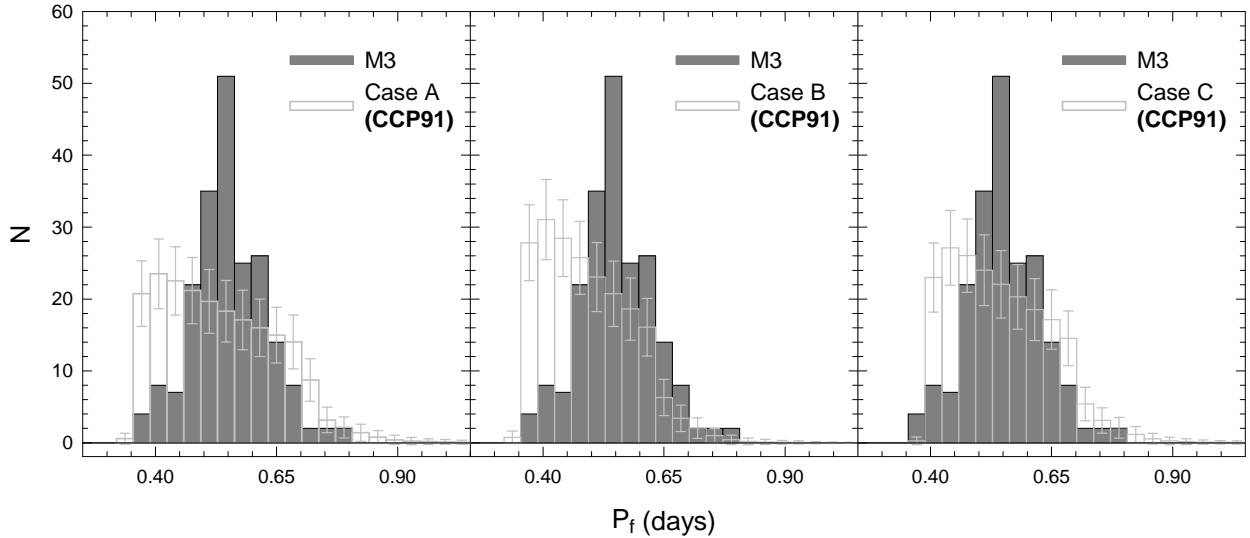,width=6.5in}}
  \caption{As in Fig.~1, but with simulations computed using the 
        Castellani et al. (1991, CCP91) evolutionary tracks. Note 
        that the results are very similar to those shown in Fig.~1. 
      }
      \label{Fig05}
\end{figure*}

Based on these results, one might wonder whether shifting the BE of the IS to 
even lower temperatures in order to further deplete the short-period end of the 
distribution, while at the same time decreasing the IS width even further in 
order to avoid an exacerbation of the discrepancy between predicted and observed 
$\sigma_{P_{\rm f}}$ values, could not bring about an additional improvement 
in the situation. To test this possibility, we have pushed both the BE 
temperature and IS width to what might arguably be their lowest possible 
bounds, computing thus a set of 5000 simulations 
for a BE that is cooler than in Case~A by 300~K, and an IS width 
$\Delta\log T_{\rm eff}^{\rm IS} = 0.065$. We will refer to this rather 
extreme combination as ``Case~D'' in what follows. 

The co-added Case~D simulations are compared against the M3 distribution in 
Fig.~4. One now finds that the short-period end of the distribution has been 
somewhat depleted, the minimum period been noticeably longer than had been 
found heretofore. However, there still remains a predicted excess of 
short-period stars with periods $P_{\rm f} \lesssim 0.45$~d, as well as an 
excess of longer-period RRL with periods $P_{\rm f} \gtrsim 0.65$~d. 

As a consequence of the smaller number of short-period stars, the statistical 
tests for Case~D do reveal some improvement with respect to the previous 
cases. The corresponding figures for this new case are as follows: 
$\langle P_{\rm KS}\rangle \simeq 2.3\%$ 
(with a maximum $P_{\rm KS} = 42.4\%$) and 
$\langle \sigma_{P_{\rm f}} \rangle = 0.097\pm 0.005$~d. 
However, one should note that still only 4.7\% of the Case~D simulations 
are found with $P_{\rm KS} > 10\%$. Moreover, the computed 
$\langle \sigma_{P_{\rm f}} \rangle$ 
value is still inconsistent with the observed one, at the $4\sigma$ level; 
in terms of Fig.~4, this is clearly indicated by the remaining excess of 
both short- and longer-period RRL stars compared to the observations.  
Hence, in spite of a rather extreme combination of IS width and 
placement which should have minimized the discrepancy between models and 
observations, Case~D is also unable to satisfactorily account for the M3 
period distribution. 

Is it possible that the noted discrepancy could be 
traced to the specific set of evolutionary tracks employed? In order to test 
this possibility, we have also computed HB simulations following a procedure 
similar to that described in \S\S2-5, but using instead the evolutionary tracks 
computed by Castellani, Chieffi, \& Pulone (1991). We again computed sets of 
5000 HB simulations for each of Cases~A, B, and C. We find that the results 
depend but slightly on the actual set of HB tracks employed. For Case~A, we 
find $\langle P_{\rm KS}\rangle \approx 1.7 \times 10^{-4}$, with a highest 
value of $P_{\rm KS} \simeq 7.6\%$; for Case~B, we find 
$\langle P_{\rm KS}\rangle \approx 2.5 \times 10^{-9}$, with a highest value 
of $P_{\rm KS} \simeq 1.1 \times 10^{-6}$;
finally, for Case~C we obtain 
$\langle P_{\rm KS}\rangle \approx 0.66\%$, with a highest value 
of $P_{\rm KS} \simeq 17.9\%$ (but only about 0.5\% of all simulations 
presenting $P_{\rm KS} \geq 10\%$). The standard deviations of the computed 
period distributions are very similar to the values obtained using the 
original set of evolutionary tracks, being only slightly larger. 
The co-added, normalized simulations for these 
three cases are shown in Fig.~5, which should be directly compared against 
Fig.~1. It is clear that the histograms for the same cases considered in 
Fig.~1 and Fig.~5 are very similar, only careful scrutiny 
revealing some small differences, particularly in the size of the build-up 
towards short periods. Clearly, the contrast between models and 
observations, first noted by Rood \& Crocker (1989), cannot be simply due 
to the specific set of HB models adopted in the simulations, though further 
experiments using different sets of evolutionary tracks might also prove 
instructive. 

In addition, one might wonder whether, if the RRL luminosities predicted by 
our models are incorrect, eventual changes in the HB luminosity might not lead 
to better agreement between the predicted and observed period distributions. 
In this sense, the histograms in Fig.~1 indicate that a major problem with 
the canonical predictions is that they foresee too many short-period variables, 
compared to the observations. Given the well-known fact that the RRL IS is 
sloped in the H-R diagram, with the blue and red edges becoming cooler as the 
luminosity increases, a higher predicted luminosity for the RRL would move the 
IS BE towards lower temperatures, thus decreasing the number of expected 
short-period variables and increasing the number of expected longer-period 
RRL. 

The mean absolute magnitude of the RRL in our simulations, at $Z = 0.001$, 
is around $M_V = 0.615$~mag. Assuming, for M3, the amount of 
alpha-enhancement provided in Table~2 of Carney (1996), and using the 
prescriptions of Salaris, Chieffi, \& Straniero (1993), this corresponds 
to a ${\rm [Fe/H]} = -1.49$. Given the available calibrations of the HB 
absolute magnitude, is there much room for an increase in the HB luminosity 
over our value? 

We can take a representative calibration of the ``long'' (i.e., bright) RRL 
distance scale (Walker 1992), and evaluate what HB magnitude should be 
expected for the quoted [Fe/H] value. According to Walker's eq.~(3), we 
get $M_V = 0.507$~mag at ${\rm [Fe/H]} = -1.49$. Hence our predicted HB 
is fainter than foreseen by ``bright'' calibrations of the HB luminosity 
by 0.108~mag, or by 0.043 in $\log\,L$. Given the dependence of the slope 
of the IS BE on luminosity from several different sources (e.g., Popielski
et al. 2000; Caputo et al. 1987), we find that this has but a small impact 
on the temperature of the BE, with $\delta\log T_{\rm eff}^{\rm BE} < 0.0035$. 
This, according to the period-mean density relation, implies a shift in the 
minimum periods towards longer values by $\delta \log P \lesssim +0.012$, 
or a shift from the ``pile-up'' values in Fig.~1 from $P \approx 0.38$~d 
to $P \lesssim 0.391$~d. This will readily be seen as too little to explain 
the discrepancy between model predictions and the observations in regard to 
the existence and placement of the pile-up at short periods. On the other 
hand, one may also note that if the luminosity shift is simultaneously 
taken into account, one finds instead a shift of the pile-up period to 
$P \lesssim 0.425$~d. However, the luminosity effect operates on the whole 
of the IS, so that the conflict between models and observations that is 
apparent at the longer-period end of the distributions in Fig.~1 would become 
even more pronounced, especially for Cases~A and C. For Case~B, the change 
would effectively be similar to the one previously carried out when going 
from Case~B to Case~C. We conclude that the only way for a change in predicted 
HB luminosity to help explain the observations would be for the IS to get even 
narrower than 0.07 as the luminosity increased. However, according to Fig.~7 
in Popielski et al. (2000), the IS width actually {\em increases} with 
increasing HB luminosity.  

Note that usage of pulsation periods in these tests  
renders us immune to the intrinsic problems related to the determination 
of equilibrium colors and temperatures of RRL stars (Rood \&  
Crocker 1989; Bono, Caputo, \& Stellingwerf 1995).

\section{Breakdown of the Canonical Framework?}
M3's peaked period distribution was already obvious in  
Oosterhoff's (1939) Fig.~1, and also in Fig.~36 of Christy (1966), 
Fig.~2 of Stobie (1971), Fig.~23 of Cacciari \& Renzini (1976), 
Fig.~5b of Castellani \& Quarta (1987), and Fig.~7 of Corwin \& Carney (2001). 
Rood \& Crocker (1989) were the first 
to note that this distribution might be in conflict with the models, 
though no statistical tests were performed at the time. More than a century 
after Bailey's (1902) data were collected, and some 64 years since 
Oosterhoff used such data to produce a plot first showing the sharp peak in 
the M3 period distribution, we are now able, for the first time, to 
establish that the M3 period distribution is incompatible with 
canonical model predictions. 

\begin{figure*}[ht]
  \figurenum{6}
  \centerline{\psfig{figure=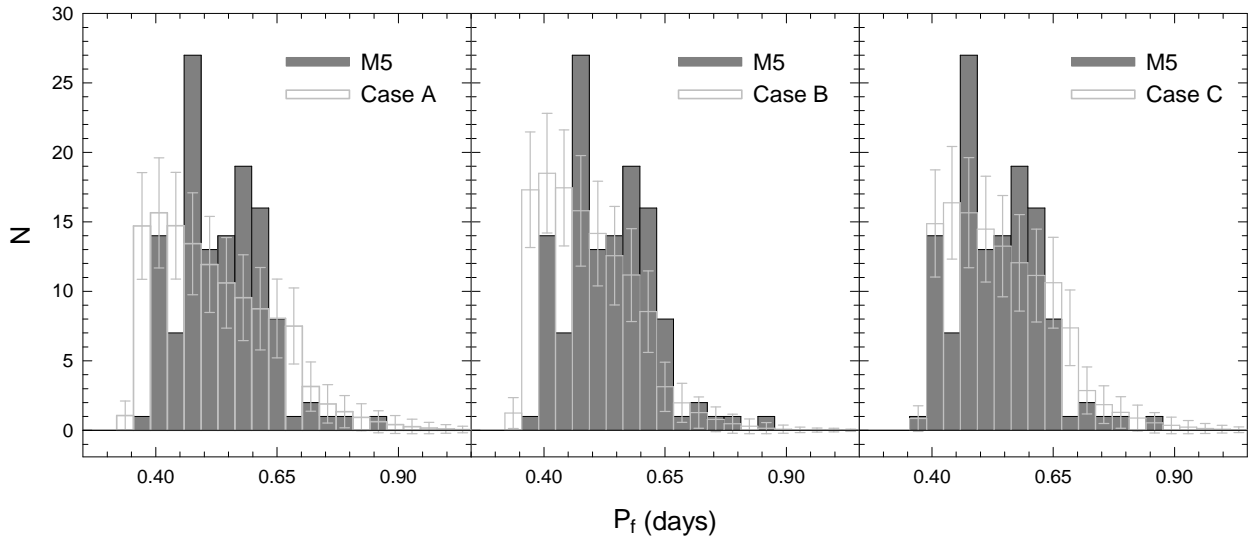,width=6.5in}}
  \caption{As in Fig.~1, but for M5.}
      \label{Fig06}
\end{figure*}

What is the reason for the discrepancy? This is unclear at present, 
but R. T. Rood (2003, priv. comm.) has hypothesized that slower evolution, 
possibly related to pulsationally-induced mass loss (Willson \& Bowen 1984), 
close to the transition region between RRab and RRc pulsators, could ``trap'' 
stars there and lead to the observed feature. Indeed, a connection between 
RRL pulsation (which is traditionally thought of as an ``envelope-only'' 
phenomenon) and interior evolution has previously been made in regard to 
period change rates (e.g., Sweigart \& Renzini 1979; Rathbun \& Smith 1997), 
but also in terms of a possible connection between mass loss, HB evolution, 
and the RRd phenomenon (Koopmann et al. 1994). While it is not clear how 
exactly the aforementioned effect might operate in the case of OoI GCs, 
the following might be a possibility for OoII GCs, which seem to present a 
similar pile-up effect among its RRc stars (see below). One might 
conjecture that the redward-evolving RRL, when 
reaching the c-ab transition line, might suffer a sudden episode of 
mass loss, which might be sufficient to momentarily drive {\em blueward} 
evolution for the star (Koopmann et al. 1994). However, as soon as the 
star evolved away from the transition line, mass loss would cease, and the 
evolution would then proceed along a similar redward path as was originally 
the case---until the transition line is reached again. While the connection 
with pulsation is speculative, a similar kind of ``evolutionary trapping'' 
related to mass loss was indeed found in the detailed evolutionary 
computations by Koopmann and co-authors. Further calculations similar to 
those carried out by Koopmann et al. would certainly prove worthwhile. 
It might also be interesting to study the role of rotation in this 
regard, particularly given its suggested connection with the Blazhko 
effect (see, e.g., Smith et al. 2003 and references therein)---which, as 
we shall see, may also contribute to the pile-up effect in M3. 

Inspection of the M3 CMD at the RRL region (Bakos \& Jurcsik 2000; Corwin \& 
Carney 2001) does reveal that the ab region is much 
more densely populated than the RRc region. This 
is in sharp contrast with the canonical model predictions, in 
which the RRL strip is fairly uniformly populated from blue to 
red (Fig.~3), 
and the larger number of RRab variables compared to the RRc ones in OoI 
globulars is primarily due to evolutionary hysteresis (van Albada \& Baker 1973), 
not to a ``pile-up'' of stars in the RRab region (see, e.g., Fig.~3 in Caputo, 
Castellani, \& Tornamb\`e 1978). In this regard, it is worth noting that 
the hysteresis mechanism implies a large overlap in temperatures and colors 
between RRab and RRc pulsators (Catelan 1993), which is {\em not} 
observed in M3 (Bakos \& Jurcsik 2000; Corwin \& Carney 2001). 

In order to provide an ad-hoc explanation for 
the color distribution of M3 RRL's within the scope of standard evolutionary 
models, we would be forced to invoke a multimodal mass distribution on the 
ZAHB (Rood \& Crocker 1989). 
One possibility that immediately comes to mind is to invoke one sharply peaked 
mass mode lying inside the RRab region, another primarily responsible for the 
blue HB and RRc components, and a third one accounting for the bulk of the 
red HB stars. However, it is unlikely that such an ad-hoc solution would prove 
satisfactory, given the fact that HB evolutionary tracks for the relevant 
metallicity show pronounced ``blue loops'' away from the ZAHB: 
if all stars started their HB evolution at the same ZAHB location, a 
significant spread in temperatures (hence periods) should still be expected, 
as is particularly evident from Fig.~3 in Caputo et al. (1978) and 
Figs.~6b and 9b from Catelan et al. (2001). 

Given the presence of these blue loops, one might then speculate that the 
ZAHB mass distribution peak responsible for the observed pile-up should
actually be placed just to the red 
side of the IS, wherefrom a blue loop just long enough to reach the pile-up 
region would originate. According to our models, and assuming hysteresis 
operates in the ``either-or'' zone, such a loop corresponds to a ZAHB mass of 
about $0.655\,M_{\odot}$ (with a ZAHB position $\log\,T_{\rm eff} \simeq 3.789$). 
In this scenario, another ZAHB mass peak would be located farther to the red 
along the red HB, and a third one on the blue HB. In this case, 
the demise of short-period RRc stars would be explained by the fact that 
i)~the region of the BE of the IS would correspond to the tail of this 
mass mode; ii)~the blue HB stars would not spend enough time inside the RRL 
strip as they evolve redward to the asymptotic giant branch. The one problem 
with this ad-hoc explanation is that it fails to explain the existence of 
sharp peaks in the period distributions of OoII GCs as well, as we will see 
below, at least under the canonical framework---which invokes redward 
evolution for the RRL stars, and most certainly also red HB stars, in OoII 
globulars. 

Another effect which may be of importance in interpreting Fig.~1  
is provided by Fig.~1 in Bakos \& Jurcsik (2000), which suggests that 
the RRab variables presenting the Blazhko effect are more clumped in  
$M_V$ than their non-Blazhko counterparts. We  
confirm that the Blazhko RRab's do have narrower $M_V$, color, and period 
distributions, compared to the non-Blazhko RRab variables. The 
corresponding standard deviations of the distributions, for the non-Blazhko 
RRab's, are $\sigma_V = 0.085$~mag, $\sigma_{\bv} = 0.034$~mag, 
$\sigma_{P_{\rm f}} = 0.064$~d; whereas for the Blazhko variables we find 
instead $\sigma_V = 0.069$~mag, $\sigma_{\bv} = 0.030$~mag, 
$\sigma_{P_{\rm f}} = 0.049$~d. However, we caution that, in spite of the 
consistently smaller standard deviations in the Blazhko case, two-sample K-S 
tests are not able to confirm that the Blazhko and non-Blazhko RRL 
have significantly different distributions in any of these 
parameters. Therefore, the suggested difference should 
be subject to further tests before it can be considered real. 

Is the peaked distribution in fundamentalized periods a universal 
characteristic of OoI systems, or does it instead reveal a problem that 
somehow applies exclusively to M3? Fig.~1 in Oosterhoff (1939) already 
suggested that the M5 
period distribution is flatter than M3's---thus indicating that the latter's 
sharply peaked distribution is likely not due to the existence of 
a very numerous, but yet undetected, population of low-amplitude RRL 
stars. In Fig.~6 we provide a histogram of fundamentalized 
periods for the RRL variables in M5, similar to what was done in Fig.~1 
for M3. The plot is based on data provided in the Clement et al. (2001)
online catalogue, and implies a 
$\sigma_{P_{\rm f}} = 0.088$~d. This is indeed larger than 
the value derived for M3, which is consistent with our interpretation that 
M3's small $\sigma_{P_{\rm f}}$ is due in part to a relatively 
unpopulated RRc region: the fraction of RRc stars is higher in M5 than in M3, 
with approximately 30\% of the M5 RRL's pulsating in the first overtone mode.

\begin{figure}[t]
  \figurenum{7}
  \centerline{\psfig{figure=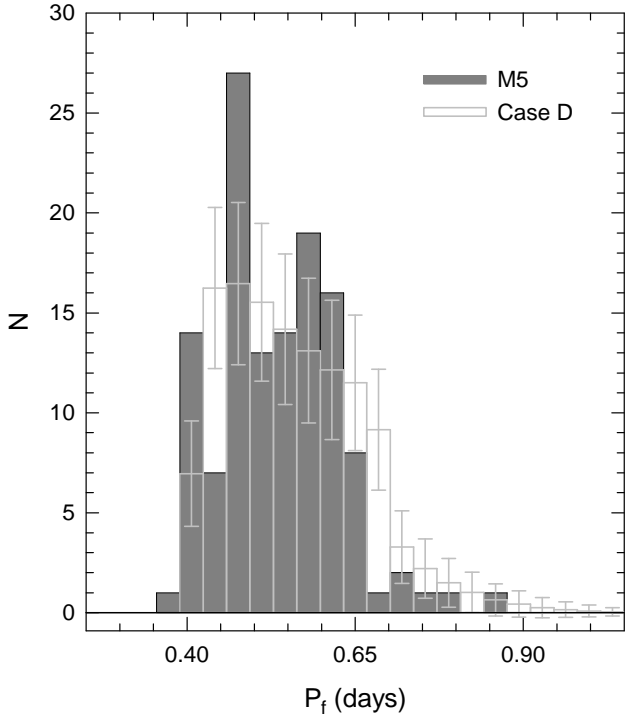,width=3.25in}}
  \caption{As in Fig.~4, but for M5. The statistics reveals this case 
    to be less successful than Case~C at reproducing the M5 period 
    distribution, contrary to what happens in the case of M3.  
      }
      \label{Fig07}
\end{figure}

Are models specifically computed for M5 
able to reproduce its period distribution? To investigate 
this, we have computed additional sets of 5000 HB simulations 
with input parameters similar to those described for M5 in Catelan (2000), 
and analogous to Cases~A, B, C for M3 (\S\S5,6). Those were compared with 
the observed period distribution using a two-sample K-S test. For Case~A, 
we find  a mean probability of 
$\approx 5\%$ that the observed and computed distributions are drawn 
from the same parent distribution. The models also give 
$\langle \sigma_{P_{\rm f}}\rangle = 0.121 \pm 0.008$~d---which is 
inconsistent with the observed value at the $4.3\sigma$ level. Case~B 
models for M5, in turn, give  
$\langle P_{\rm KS}\rangle \approx 5.6\times 10^{-3}$ and 
$\langle \sigma_{P_{\rm f}}\rangle = 0.094 \pm 0.007$~d. As for M3, the K-S 
test shows worse agreement between models and observations for a smaller IS 
width, although the smaller computed $\sigma_{P_{\rm f}}$
is now consistent with the observed one. Finally, 
Case~C models for M5 yield $\langle P_{\rm KS}\rangle \approx 27\%$ and 
$\langle \sigma_{P_{\rm f}}\rangle = 0.106 \pm 0.008$~d. While the standard 
deviation differs from the observed value at the $2.2\sigma$ level, 
the K-S test now indicates that some satisfactory matches between models and 
observations can be achieved for M5 in this case, with the maximum $P_{\rm KS}$ 
found in the Case~C simulations being 98.7\% and 15.7\% of them  
being found with $P_{\rm KS} \geq 50\%$. Detailed comparisons between models and 
observations for M5 are presented in Fig.~6. We note, in addition, that while 
the Case~D simulations improved the agreement between observed and predicted 
period distributions somewhat in the case of M3, the same does not happen 
in the case of M5, for which the agreement deteriorates instead: for Case~D, 
we find $\langle P_{\rm KS}\rangle \approx 12.1\%$ (with a maximum 
$P_{\rm KS} = 76.9\%$, but only 0.66\% of the simulations being found 
with $P_{\rm KS} \geq 50\%$), and 
$\langle \sigma_{P_{\rm f}}\rangle = 0.103 \pm 0.008$~d. Comparison between 
the 5000 co-added Case~D simulations and the M5 observations is provided in 
Fig.~7. 

The somewhat better agreement between models and observations, in 
the case of M5, is due to the presence of a more substantial population of 
short-period RRc variables in this cluster compared to M3 (in spite of a 
substantially larger overall RRL population in the latter), along 
with a less pronounced peak in the M5 period distribution. Additional 
studies of the M5 RRL population based on the image-subtraction package 
{\sc isis} (Alard 2000) might prove very important in defining the relative 
proportion of short-period, low-amplitude RRc's (and RRab's) in this cluster, 
and in conclusively establishing whether pronounced peaks in the RRL period 
distribution similar to M3's might be present in this cluster as well. In 
this sense, it should also be extremely important to establish whether the 
population of Blazhko variables in M5 is indeed a mere two or three stars, 
as currently indicated by the Clement et al. (2001) catalogue, because  
the peaked period distribution in M3, as we have seen, seems to receive 
a significant contribution from the Blazhko variables that are present 
in that cluster.  

We can also check to see whether the noted anomaly is a characteristic solely 
of OoI systems, or whether instead it also affects OoII globulars. In Fig.~8, 
we show the fundamentalized period distributions for RRL variables in the 
three OoII GCs with the largest number of RRL according to the Clement et al. 
(2001) catalogue, namely: M15 (NGC~7078), M53 (NGC~5024), and M68 
(NGC~4590). The period values were taken directly from the Clement et al. 
entries; in the case of M68, we adopted the entries corresponding to the 
Walker (1994) study. 
Due to the smaller number of variables 
in OoII GCs, a smaller number of bins was used in this case, though again we 
find our results, qualitatively, to be insensitive to the specific choice of 
bin number and size.   
As in the cases of M3 and M5 previously discussed, we have 
fundamentalized the periods of the RRc's by adding 0.128 to their log-periods. 
A similar procedure was applied to any double-mode (RRd) or (candidate) 
second-overtone (RRe) pulsator that might be present. Note that the fraction 
of RRd plus RRe variables, compared to the total numbers of RRc, RRd, and RRe
pulsators, is small in M53 (6.7\%) and M15 (14.5\%), but not so in 
M68 (43\%). These fractions are even smaller in the cases of M3 and M5; in 
the latter, there is not a single RRd variable according to the Clement et al. 
catalogue, and but one RRe candidate; whereas in the former, it appears that 
fewer than 4\% of the RRL are RRd or (candidate) RRe pulsators. 

Unfortunately, 
detailed statistical tests are not as conclusive in the case of OoII GCs, 
given the intrinsically smaller number of RRL compared to the case of M3, 
or even M5. Moreover, it has been argued in the literature that, whereas 
there are indications that the RRL variables in OoII systems are indeed 
evolved away from a position on the blue ZAHB, canonical models fail to 
produce such evolved RRL in adequate numbers (Pritzl et al. 2002 and 
references therein). Therefore, until the evolutionary status of the RRL 
in OoII GCs is adequately established, a detailed statistical comparison 
between the predictions of canonical HB models and the observed period 
distributions in OoII GCs may not prove conclusive. For this reason, we defer 
such a detailed comparison between models and observations for OoII GCs 
to a future occasion.

\begin{figure*}[ht]
  \figurenum{8}
  \centerline{\psfig{figure=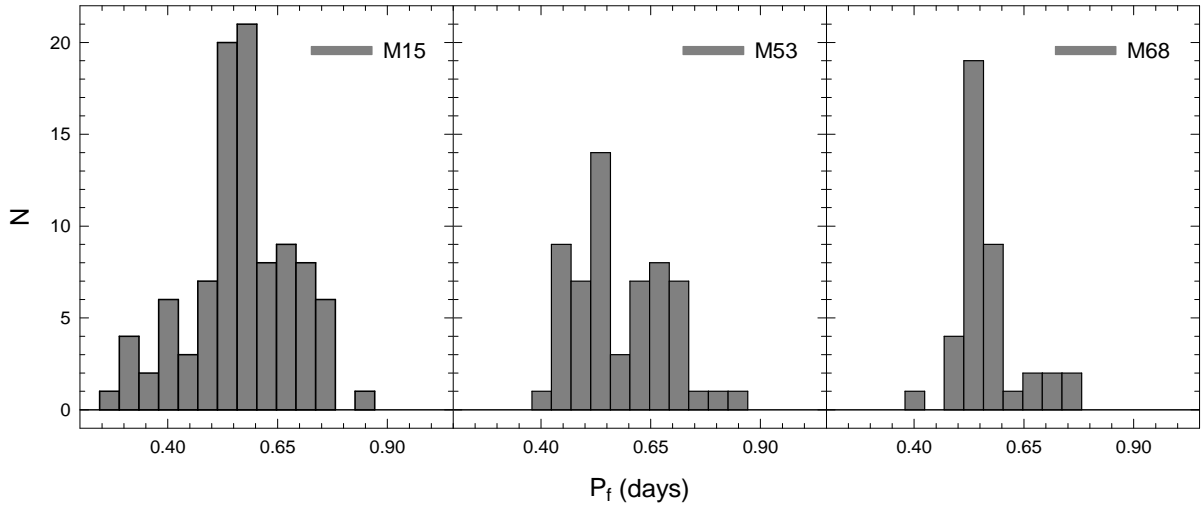,width=6.25in}}
  \caption{Fundamentalized period distributions for the RRL in well-populated  
        OoII GCs: M15 (left), M53 (center), and M68 (right).  
      }
      \label{Fig08}
\end{figure*}

Qualitatively though, 
one may note, from Fig.~8, that at least M15 and M68 do show signs of having 
a sharp peak in their period distributions, whereas the same is not as 
obvious in the case of M53, in spite of its larger number of RRL variables
(compared to M68). Interestingly, the peak in the period distribution is 
located at a similar (perhaps slightly shorter) period in comparison with 
the case of M3. Moreover, there is again no sign of a pile-up at the shorter 
periods, even though the pile-up effect might perhaps be expected to be strong 
in OoII globulars as well, if RRL variables are evolved away from a 
position on the blue ZAHB, since in this case (redward) 
evolution would be slower close to the BE of the IS than close to 
the IS RE (Catelan 1994; see also Figs.~3d,e in Caputo et al. 1978). 
The standard deviations of the period 
distributions, in the cases of M15 and M53---$\sigma_{P_{\rm f}} = 0.117$~d 
and 0.107~d, respectively---are consistently larger than found previously 
for M3 and M5. On the other hand, the M68 distribution, with 
$\sigma_{P_{\rm f}} = 0.074$~d, seems even narrower than is the case for the 
OoI GCs.

In the spirit of Rood's speculative scenario for the pile-up of stars on the 
RRab side of the IS in OoI systems, one may hypothesize that the same is 
happening in at least some OoII GCs, but {\em on the opposite side of the
IS}---that is, on the RRc side. Hence, in OoI systems, the RRL might get 
``trapped'' just to the red of the RRab-RRc transition line, while they 
are evolving blueward; whereas, in OoII systems, the RRL might instead get 
``trapped'' just to the blue of this transition line, in the course of 
their redward evolution. If so, we would expect to find evidence, in OoII 
GC CMDs and Bailey diagrams, of a pile-up of RRc variables close to the 
transition line. 

In this regard, we first note that the peaks in the M15 and M68 
period distributions are 
indeed primarily due to the RRc variables; the case of M68 is particularly 
noteworthy. This cluster has 15 RRc stars, 14 of which have periods in the 
range 0.34 to 0.39~d, corresponding to fundamentalized periods 0.47 to 0.52~d. 
The single exception is V5, with a period of 0.282~d. Perhaps not surprisingly, 
we do find most of these RRc's located close to the transition line between 
RRc and RRab variables---see, for example, Fig.~13 in Walker (1994). In 
the case of M15, the lack of a CCD study of its variables based on 
state-of-the-art photometric data obtained under good seeing conditions 
limits our current possibilities; however, interesting indications are 
already provided by the Bingham et al. (1984) photographic study, as well 
as by the $V$, $R$ data obtained by Silbermann \& Smith (1995). In the CMD 
showing the detailed topology of the IS from Bingham et al. (see their Fig.~14), 
one again encounters a clear pile-up of the RRc variables close to the c-ab 
transition line. That these stars are 
primarily responsible for the peak in the period distribution is particularly 
evident by confronting the quoted figure with Fig.~18 in the same paper; it 
is also seen that the RRd's contribute to the effect as well, being situated 
just to the red (long-period side) of the peak in the period distribution.  
Importantly, one again finds a demise of RRc variables close to 
the BE of the IS, contrary to what is predicted by the simulations: it is 
difficult to see, in the canonical evolutionary scenario, how the middle 
region of the IS could be more heavily populated than the vicinity of the IS 
BE. These conclusions are also consistent with the Silbermann \& Smith CMD 
(their Fig.~11), Bailey (Fig.~13), and period-color diagrams (Fig.~14).  
To the best of our knowledge, similar diagrams are currently lacking 
for M53. 

Perhaps noteworthy as well is the suggested presence, in M15 and M53 alike 
(but not in M68), of a secondary ``hump'' in the period distributions, at 
around $P_{\rm f} \approx 0.67$~d; a similar feature can be noted to be present 
in the cases of M3 and M5, though at a slightly shorter period---namely, 
$P_{\rm f} \approx 0.62$~d. Such a feature is not generally present in the 
computed model distributions (see the co-added, normalized results in 
Figs.~1 through 7). Along with the lack of substantial populations of 
short-period RRc variables located close to the BE of the IS in 
the clusters that we have studied---which may actually constitute a fairly  
universal feature, judging from the plots in Cacciari \& Renzini (1976) 
and Castellani \& Quarta (1987), M5 possibly representing an exception 
to this rule---besides the sharp period distributions in 
many of them, this seems to point to a serious challenge to our current 
understanding of the interplay between pulsation and evolution of RRL 
stars in GCs, and thereby of the Oosterhoff dichotomy itself. The yellow 
flag raised by Rood \& Crocker back in 1989 is now looking quite red indeed.

\acknowledgments It is a pleasure to thank R. T. Rood, J. Jurcsik, A. W. Stephens, 
A. V. Sweigart, and D. Castillo for interesting discussions and comments. I thank 
F. R. Ferraro and D. A. VandenBerg for providing data in advance of publication. 
I am grateful, in particular, to H. A. Smith for many useful discussions and 
insightful comments, and for critically reading, and commenting upon, several 
drafts of this paper. Some comments by an anonymous referee were helpful in 
revising the manuscript. This work was supported by Proyecto FONDECYT Regular 
No. 1030954.

\vfill\eject

\end{document}